\documentclass[aps,twocolumn,showpacs,superscriptaddress]{revtex4}
\usepackage{epsfig}
\newcommand{\ben}{\begin{eqnarray}}
\newcommand{\een}{\end{eqnarray}}

\newcommand{\bef}{\begin{figure}[h!bt]\centering}
\newcommand{\eef}{\end{figure}}

\begin{document}
\title{Test the time-reversal modified universality of the Sivers function}
\author{Zhong-Bo Kang}
\email{kangzb@iastate.edu}
\affiliation{Department of Physics and Astronomy,
                 Iowa State University,
                 Ames, IA 50011, USA}
\author{Jian-Wei Qiu}
\email{jwq@iastate.edu}
\affiliation{Department of Physics and Astronomy,
                 Iowa State University,
                 Ames, IA 50011, USA}
                 
\begin{abstract}
We derive the time-reversal modified universality for both quark and gluon Sivers function from the parity and time-reversal invariance of QCD.  We calculate the single transverse-spin asymmetry of inclusive lepton from the decay of $W$ bosons in polarized proton-proton collision at RHIC, in terms of the Sivers function.  We find that although the asymmetry is diluted from the $W$ decay, the lepton asymmetry is at the level of several percent and is measurable for a good range of lepton rapidity at RHIC.  We argue that this measurable lepton asymmetry at RHIC is an excellent observable for testing the time-reversal modified universality of the Sivers function.  
\end{abstract}
\pacs{12.38.Bx, 12.39.St, 13.85.Qk, 13.88+e}
\date{\today}
\maketitle

{\it I.\ Introduction. }\
Much of the predictive power of perturbative Quantum Chromodynamics (QCD) 
is contained in factorization theorems \cite{CSS-fac}.  They normally include two assertions.  One is that a physically measured quantity can be factorized into some perturbatively calculable short-distance hard parts convoluted with nonperturbative long-distance distribution functions (or matrix elements).  The other is the {\em universality} of the nonperturbative distribution functions.  Predictions follow when processes with different hard scatterings but the same distribution functions are compared.  With one set of universal parton distribution functions (PDFs) 
the leading power collinear QCD factorization formalisms have been very successful in interpreting and predicting almost all existing data 
from high energy collisions with 
momentum transfers larger than a few GeV \cite{PDFs}.

The phenomenon of single transverse-spin asymmetry (SSA),
$A_N \equiv (\sigma(\vec{S}_\perp)-\sigma(-\vec{S}_\perp))
/(\sigma(\vec{S}_\perp)+\sigma(-\vec{S}_\perp))$, 
defined as the ratio of the difference and the sum of the 
cross sections when the single transverse spin vector $\vec{S}_\perp$ 
is flipped, was first observed in the hadronic $\Lambda^0$ production 
at Fermilab in 1976 as a surprise \cite{Bunce:1976yb}.  
Large SSAs, as large as 30 percent, have been consistently observed in various experiments involving one polarized hadron at different collision energies \cite{ssa_review}.  
The size of the SSAs presented a challenge to the leading power collinear QCD factorization formalism \cite{Kane:1978nd}, and provided 
an excellent opportunity to probe a new domain of QCD dynamics.

Two widely discussed theoretical approaches 
have been proposed to evaluate the observed SSAs in QCD.
One generalizes the QCD collinear factorization approach to the next-to-leading power in the momentum transfer \cite{QS_fac}, and attributes the SSA to the quantum interference of scattering amplitudes with different numbers of active partons \cite{Efremov,qiu}.  The size of the asymmetry is determined by new three-parton correlation functions \cite{Kang:2008ey}. 
This generalized collinear factorization approach is more relevant 
for the SSAs of cross sections whose momentum transfers
$Q\gg \Lambda_{\rm QCD}$.  
The other approach factorizes the cross sections so as the SSAs 
in terms of the transverse momentum dependent (TMD) parton distributions
 \cite{Collins:2002kn,Boer:2003cm,Ji:2004xq,Collins:2004nx,Collins:2007nk},
and attributes the SSAs to the Sivers function \cite{Sivers} 
(or the Collins function if a final-state hadron was observed \cite{Collins}).  
The TMD factorization approach
is more suitable for the SSAs of cross sections 
with two very different momentum transfer scales, 
$Q_1\gg Q_2 \gtrsim \Lambda_{\rm QCD}$.  
These two approaches each have their kinematic domain of validity,
they were shown to be consistent with each other in the kinematic
regime where they both apply \cite{UnifySSA}.

However, there is one crucial difference between these two factorization approaches besides the difference in kinematic regimes where they apply.  The Sivers function in the TMD factorization approach could be process dependent, while all distribution functions in the collinear factorization approach are universal. It was predicted by Collins \cite{Collins:2002kn}
on the basis of time-reversal arguments that the quark Sivers function in semi-inclusive deep inelastic scattering (SIDIS) and in Drell-Yan process (DY) have the same functional form but an {\em opposite sign}, a time-reversal modified universality.  In this Letter, we derive the same time-reversal modified universality for both quark and gluon Sivers function from the parity and time-reversal invariance of QCD.

The experimental check of this time-reversal modified universality of the  Sivers function would provide a critical test for the TMD factorization 
approach 
\cite{Collins:2002kn,Boer:2003cm,Ji:2004xq,Collins:2004nx,Collins:2007nk}.  
Recently, the quark Sivers function has 
been extracted from data of SIDIS experiments \cite{Anselmino:2008sga}.  
Future measurements of the SSAs in DY production have been planned
\cite{RHIC-dy}.  In this Letter, we present our calculation of the SSAs of inclusive single lepton production from the decay of $W$ bosons. The $W$ production and DY share the same Sivers function.  We find that although the asymmetry is diluted from the decay of $W$ bosons, the lepton asymmetry is  significant and measurable 
for a good range of lepton rapidity at RHIC. 
We show that the lepton SSAs provide the
better flavor separation of the quark Sivers function than what 
the standard DY can do.
We also show that the lepton SSAs are sharply peaked at 
transverse momentum $p_T\sim M_W/2$ with $W$ mass $M_W$.  
Since leptons from heavy quarkonium decay and other potential backgrounds are unlikely to be peaked at the 
$p_T\sim M_W/2$, we argue that the SSA of inclusive high $p_T$ leptons at RHIC is an excellent observable for testing the time-reversal modified universality of the Sivers function.  

{\it II.\ The QCD prediction. }\
The predictive power of the TMD factorization approach to the SSAs 
relies on the universality of the TMD parton distributions.  
For the lepton-hadron SIDIS, 
$\ell(l)+h(p,\vec{S})\to \ell'(l')+h'(p')+X$,
the factorized TMD quark distribution has the following gauge
invariant operator definition \cite{tmd_gauge},
\ben
f_{q/h^\uparrow}^{\rm SIDIS}
(x,\mathbf{k}_\perp,\vec{S})
&=& 
\int \frac{dy^- d^2\mathbf{y}_\perp}{(2\pi)^3}\,
e^{ixp^+ y^- - i\,\mathbf{k}_\perp\cdot \mathbf{y}_\perp}
\nonumber\\
& \times &
\langle p,\vec{S}|
\overline{\psi}(0^-,\mathbf{0}_\perp)
\Phi_n^\dagger(\{\infty,0\},\mathbf{0}_\perp)
\nonumber\\
& \times &
\Phi_{\mathbf{n}_\perp}^\dagger
     (\infty,\{\mathbf{y}_\perp,\mathbf{0}_\perp\})
\frac{\gamma^+}{2}
\label{qkt_dis}\\
& \times &
\Phi_n(\{\infty,y^-\},\mathbf{y}_\perp)
\psi(y^-,\mathbf{y}_\perp)
|p,\vec{S}\rangle ,
\nonumber
\een
where $y^+=0^+$ dependence is suppressed
and the gauge links from the final-state interaction 
of SIDIS are
\ben
\Phi_n(\{\infty,y^-\},\mathbf{y}_\perp) 
&\equiv &
{\cal P}e^{-ig\int_{y^-}^{\infty} dy_1^- 
              n^\mu A_\mu(y_1^-,\,\mathbf{y}_\perp)}\, ,
\nonumber\\
\Phi_{\mathbf{n}_\perp}(\infty,\{\mathbf{y}_\perp,\mathbf{0}_\perp\}) 
&\equiv &
{\cal P}e^{-ig\int_{\mathbf{0}_\perp}^{\mathbf{y}_\perp} 
             d\mathbf{y}'_\perp 
            \mathbf{n}_\perp^\mu A_\mu(\infty,\,\mathbf{y}'_\perp)}\, ,
\label{g_link}
\een
where ${\cal P}$ indicates the path ordering and 
the direction $\mathbf{n}_\perp$ is pointed from 
$\mathbf{0}_\perp$ to $\mathbf{y}_\perp$. 
Here we define the light-cone vectors, 
$n^\mu=(n^+,n^-,\mathbf{n}_\perp)=(0,1,\mathbf{0}_\perp)$
and $\bar{n}^\mu=(1,0,\mathbf{0}_\perp)$, which project out the 
light-cone components of any four-vector $V^\mu$ as $V\cdot n=V^+$
and $V\cdot\bar{n}=V^-$.

For the DY, 
$h(p,\vec{S})+h'(p')\to \gamma^*(Q)[\to \ell^+\ell^-]+X$, 
the factorized TMD quark distribution is given by
\ben
f_{q/h^\uparrow}^{\rm DY}
(x,\mathbf{k}_\perp,\vec{S})
&=& 
\int \frac{dy^- d^2\mathbf{y}_\perp}{(2\pi)^3}\,
e^{ixp^+ y^- - i\,\mathbf{k}_\perp\cdot \mathbf{y}_\perp}
\nonumber\\
& \times &
\langle p,\vec{S}|
\overline{\psi}(0^-,\mathbf{0}_\perp)
\Phi_n^\dagger(\{-\infty,0\},\mathbf{0}_\perp)
\nonumber\\
& \times &
\Phi_{\mathbf{n}_\perp}^\dagger
      (-\infty,\{\mathbf{y}_\perp,\mathbf{0}_\perp\})
\frac{\gamma^+}{2}
\label{qkt_dy}\\
& \times &
\Phi_n(\{-\infty,y^-\},\mathbf{y}_\perp)
\psi(y^-,\mathbf{y}_\perp)
|p,\vec{S}\rangle
\nonumber
\een
where the past pointing gauge links were caused by the initial-state 
interactions of DY production \cite{Collins:2002kn}.  
From Eqs.~(\ref{qkt_dis}) and 
(\ref{qkt_dy}), it is easy to show that
the collinear quark distributions are process independent, 
\ben
\int d^2\mathbf{k}_\perp 
f_{q/h^\uparrow}^{\rm SIDIS}(x,\mathbf{k}_\perp,\vec{S})
=\int d^2\mathbf{k}_\perp 
f_{q/h^\uparrow}^{\rm DY}(x,\mathbf{k}_\perp,\vec{S}),
\label{quark-dis}
\een
if the same renormalization scheme was used for the
ultraviolet divergence of the $\mathbf{k}_\perp$ integration.

Let $|\alpha\rangle = |p,\vec{S}\rangle$
and $\langle \beta|$ be equal to the rest of the matrix element
in Eq.~(\ref{qkt_dis}) \cite{Kang:2008ey}.  
From the parity and time-reversal 
invariance of QCD,
$\langle \alpha_P|\beta_P \rangle=
 \langle \alpha|\beta\rangle$ and 
 $\langle \beta_T|\alpha_T \rangle=
 \langle \alpha|\beta\rangle$, where
 $|\alpha_P\rangle$ and $|\beta_P\rangle$, and
 $|\alpha_T\rangle$ and $|\beta_T\rangle$ are the parity and 
 time-reversal transformed states from the states $|\alpha\rangle$
 and $|\beta\rangle$, 
 respectively, we derive 
\ben
f_{q/h^\uparrow}^{\rm SIDIS}
(x,\mathbf{k}_\perp,\vec{S})
=
f_{q/h^\uparrow}^{\rm DY}
(x,\mathbf{k}_\perp,-\vec{S})\, .
\label{pt_inv}
\een
and conclude that 
the spin-averaged TMD quark distributions are process independent.  Following the notation of Ref.~\cite{Anselmino:2008sga}, 
we expand the TMD quark distribution as
\ben
f_{q/h^\uparrow}(x,\mathbf{k}_{\perp},\vec{S})
&\equiv &f_{q/h}(x,k_{\perp}) 
\nonumber \\
&+ & 
\frac{1}{2}\Delta^N f_{q/h^\uparrow}(x,k_\perp)\,
\vec{S}\cdot \left(\hat{p}\times \hat{\mathbf{k}}_\perp \right)\,
\label{TMDPDF}
\een
where $k_\perp=|\mathbf{k}_\perp|$, $\hat{p}$ and $\hat{\mathbf{k}}_\perp$ are the unit vectors of $\vec{p}$ and $\mathbf{k}_\perp$, respectively, 
$f_{q/h}(x,k_\perp)$ is the spin-averaged TMD distribution, 
and $\Delta^N f_{q/h^\uparrow}(x,k_\perp)$ is the Sivers function 
\cite{Sivers}.  Substituting Eq.~(\ref{TMDPDF}) into Eq.~(\ref{pt_inv}), we
obtain,
\ben
\Delta^N f^{\rm SIDIS}_{q/h^\uparrow}(x,k_\perp)
= - 
\Delta^N f^{\rm DY}_{q/h^\uparrow}(x,k_\perp)\, ,
\label{Sivers_sign}
\een
which confirms the Collins' prediction \cite{Collins:2002kn} 
that the Sivers function in SIDIS and in DY differ by a
sign.  

We define the gauge invariant TMD gluon distribution in SIDIS 
and in DY by replacing the quark operator 
$\overline{\psi}(\gamma^+/2)\psi$ 
in Eqs.~(\ref{qkt_dis}) and (\ref{qkt_dy})
by the gluon operator $F^{+\mu}F^{+\nu}(-g_{\mu\nu})$,
and the gauge links by those in the adjoint representation of SU(3)
color.  From the parity and time-reversal invariance 
of the matrix elements of the TMD gluon distribution, we
find, like Eq.~(\ref{pt_inv}),
\ben
f_{g/h^\uparrow}^{\rm SIDIS}
(x,\mathbf{k}_\perp,\vec{S})
=
f_{g/h^\uparrow}^{\rm DY}
(x,\mathbf{k}_\perp,-\vec{S})\, .
\label{gluon_sivers}
\een
Applying Eq.~(\ref{TMDPDF}) to the gluon TMD distribution, we derive
the same time-reversal modified universality for the gluon 
Sivers function, 
\ben
\Delta^N f^{\rm SIDIS}_{g/h^\uparrow}(x,k_\perp)
= - 
\Delta^N f^{\rm DY}_{g/h^\uparrow}(x,k_\perp)\, .
\label{gluon_sign}
\een
The sign change of the Sivers function is a property of the 
gauge invariant TMD parton distributions.

{\it III.\ Lepton SSAs from $W$ production. }\
The SSAs of $W$ production at RHIC,
$A(p_A,\vec{S}_\perp)+B(p_B)\to W^{\pm}(q) \to \ell^{\pm}(p)+X$,
were proposed in Refs.~\cite{ssa_W} to measure 
the Sivers function.  However, it is difficult to
reconstruct $W$ bosons by the current detectors at RHIC.  
We present here our predictions for the SSAs of inclusive lepton
production from the decay of $W$ bosons.  We also 
present the SSAs of $W$ production for a comparison. 

We use the TMD factorization formalism because $W$ bosons at RHIC 
are likely produced with transverse momentum 
$|\mathbf{q}_\perp| \ll M_W$.  We work in a frame in which the 
polarized hadron $A$ moves in the $+z$-direction.
We have the leading order spin-averaged $W$ cross section
\ben
\frac{d\sigma_{AB\to W}}{dy_W\,d^2\mathbf{q}_\perp}
&=&
\sigma_0 \sum_{a,b}\left|V_{ab}\right|^2
\int d^2\mathbf{k}_{a\perp} d^2\mathbf{k}_{b\perp}
f_{a/A}(x_a,{k}_{a\perp})
\nonumber \\
& & \times
f_{b/B}(x_b,{k}_{b\perp}) \,
\delta^2(\mathbf{q}_\perp-\mathbf{k}_{a\perp}-\mathbf{k}_{b\perp}),  
\label{spin_avg} 
\een
where $y_W$ is the $W$ rapidity, $\sigma_0=(\pi/3)\sqrt{2}\,{\rm G_F}M_W^2/s$ is the lowest order partonic cross section with the Fermi weak coupling constant G$_F$ and $s=(p_A+p_B)^2$, $\sum_{ab}$ runs over all light (anti)quark flavors, $V_{ab}$ are the CKM matrix elements for the weak interaction.
The parton momentum fractions in Eq.~(\ref{spin_avg}) are given by
\ben
x_a = \frac{M_W}{\sqrt{s}}\, e^{y_W}, \quad
x_b = \frac{M_W}{\sqrt{s}}\, e^{-y_W}
\label{xaxb}
\een
to the leading power in $q_\perp^2/M_W^2$.  Similarly, we have
the leading order factorized spin-dependent $W$ cross section
$\Delta\sigma(\vec{S}_\perp) 
= [\sigma(\vec{S}_\perp)-\sigma(-\vec{S}_\perp)]/2$ as
\ben
&&
\frac{d\Delta\sigma_{A^\uparrow B\to W}(\vec{S}_\perp)}
     {dy_W\,d^2\mathbf{q}_\perp}
=
\frac{\sigma_0}{2}\sum_{a,b}\left|V_{ab}\right|^2
\int d^2\mathbf{k}_{a\perp} d^2\mathbf{k}_{b\perp}\,  
\nonumber \\
&& {\hskip 0.5in} \times 
\vec{S}_\perp\cdot (\hat{p}_A\times\hat{\mathbf{k}}_{a\perp})\,
\Delta^N f^{\rm DY}_{a/A^\uparrow}(x_a,{k}_{a\perp})\,
\nonumber \\
&& {\hskip 0.5in} \times
f_{b/B}(x_b,{k}_{b\perp})\, 
\delta^2(\mathbf{q}_\perp-\mathbf{k}_{a\perp}-\mathbf{k}_{b\perp})\, .
\label{spin_dpt}
\een
The SSA of $W$ production is then defined as,
\ben
A_N^{(W)}
&\equiv & \left.
\frac{d\Delta\sigma(\vec{S}_\perp)_{A^\uparrow B\to W}}
     {dy_W\,d^2\mathbf{q}_\perp} \right/
\frac{d\sigma_{AB\to W}}
     {dy_W\,d^2\mathbf{q}_\perp}\, ,
\label{An_W_def}
\een
whose sign depends on the sign of the Sivers function and 
the direction of the spin vector $\vec{S}_\perp$.

To evaluate the SSA in Eq.~(\ref{An_W_def}),
we use the parameterization of TMD parton distributions
in Ref.~\cite{Anselmino:2008sga},
\ben
f_{q/h}(x,k_\perp) 
&=& f_q(x)\,
\frac{1}{\pi\langle k_\perp^2\rangle}\, 
e^{-k_\perp^2/\langle k_\perp^2\rangle} ,
\label{TMDavg} \\
\Delta^N f_{q/h^\uparrow}^{\rm SIDIS}(x,k_\perp)
&=&
2\,{\cal N}_q(x)\,h(k_\perp)\,
f_{q/h}(x,k_\perp) ,
\label{Sivers_f} \\
h(k_\perp)
&=&
\sqrt{2e}\, \frac{k_\perp}{M_1}\, 
e^{-k_\perp^2/M_1}
\label{Sivers_h}
\een
where $f_q(x)$ is the standard unpolarized parton distribution 
of flavor $q$, 
$\langle k_\perp^2\rangle$ and $M_1$ are fitting parameters, and
${\cal N}_q(x)$ is a fitted distribution given in 
Ref.~\cite{Anselmino:2008sga}. 
By carrying out the integration 
$d^2\mathbf{k}_{a\perp}d^2\mathbf{k}_{b\perp}$ 
in Eqs.~(\ref{spin_avg}) and 
(\ref{spin_dpt}) analytically, we obtain,
\ben
A_N^{(W)} &=& 
\vec{S}_\perp \cdot (\hat{p}_A\times \mathbf{q}_\perp)\,
\frac{2\langle k_s^2\rangle^2}
     {[\langle k_\perp^2\rangle + \langle k_s^2\rangle]^2}\,
e^{-\left[
     \frac{\langle k_\perp^2\rangle-\langle k_s^2\rangle}
          {\langle k_\perp^2\rangle + \langle k_s^2\rangle}
    \right] 
    \frac{\mathbf{q}_\perp^2}{2\langle k_\perp^2\rangle}
    }
    \nonumber\\
& \times &
\frac{\sqrt{2e}}{M_1}\,
\frac{\sum_{ab}|V_{ab}|^2\left[-{\cal N}_{a}(x_a)\right] 
      f_{a}(x_a)\,f_{b}(x_b)}
     {\sum_{ab}|V_{ab}|^2\,f_{a}(x_a)\, f_{b}(x_b)} ,
\label{An_W}
\een
where $\langle k_s^2\rangle=M_1^2\, \langle k_\perp^2\rangle
     /[M_1^2+\langle k_\perp^2\rangle]$
     and the ``$-$'' sign in front of ${\cal N}_a(x_a)$ is from 
Eq.~(\ref{Sivers_sign}).
If we choose the $\vec{S}_\perp$ along the $y$-axis as 
in Ref.~\cite{Anselmino:2008sga}, 
$\vec{S}_\perp \cdot (\hat{p}_A\times \mathbf{q}_\perp)= 
q_T \cos(\phi_W)$ with 
$q_T\equiv |\mathbf{q}_\perp|$ and
azimuthal angle $\phi_W$.
For our numerical predictions below, we choose  
$\phi_W=0$ and the GRV98LO 
parton distribution \cite{GRV98} for $f_q(x)$ 
to be consistent with the usage of 
the TMD distributions of Ref.~\cite{Anselmino:2008sga}.

\bef
\psfig{file=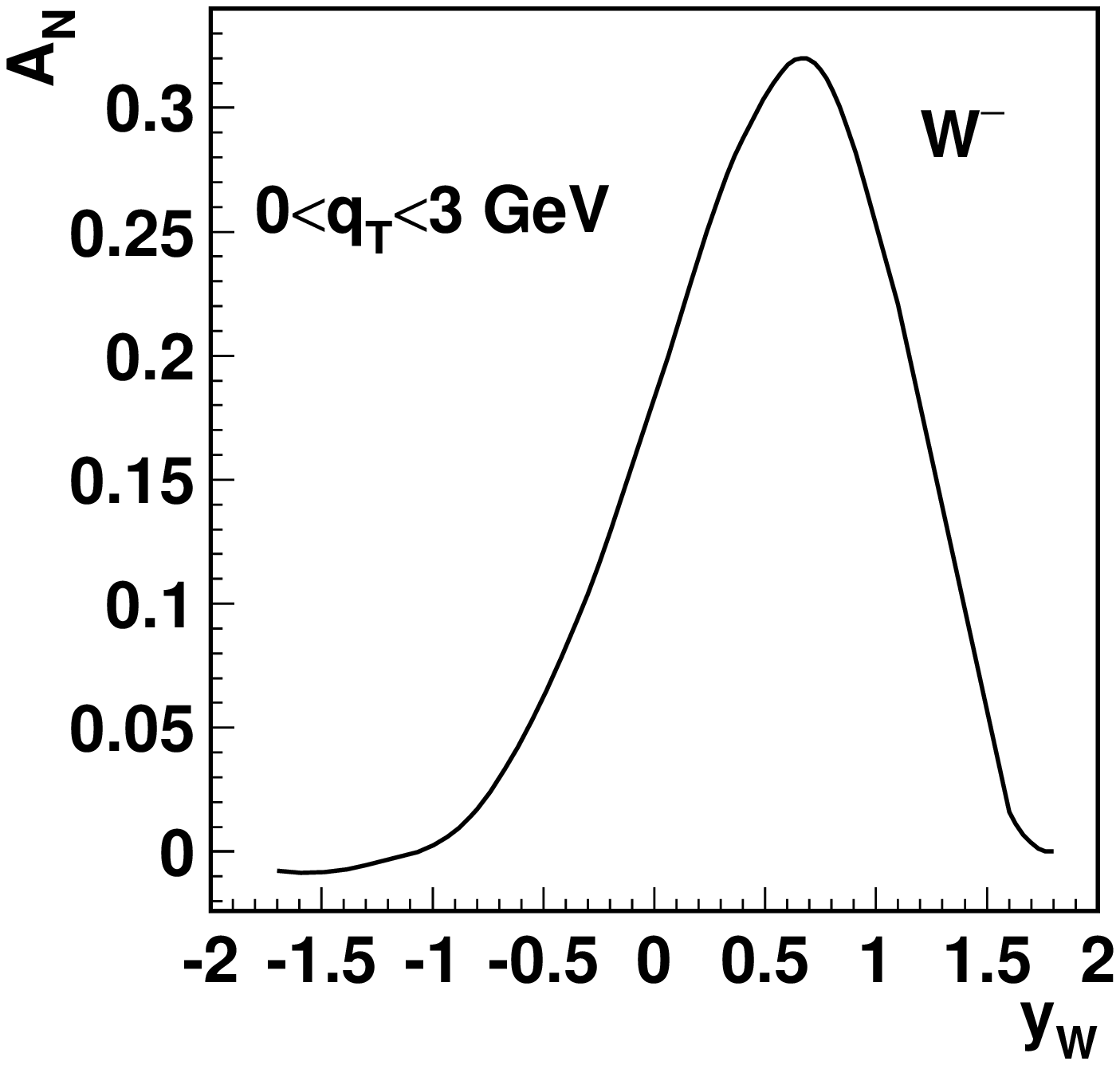,width=1.55in}
\hskip 0.1in
\psfig{file=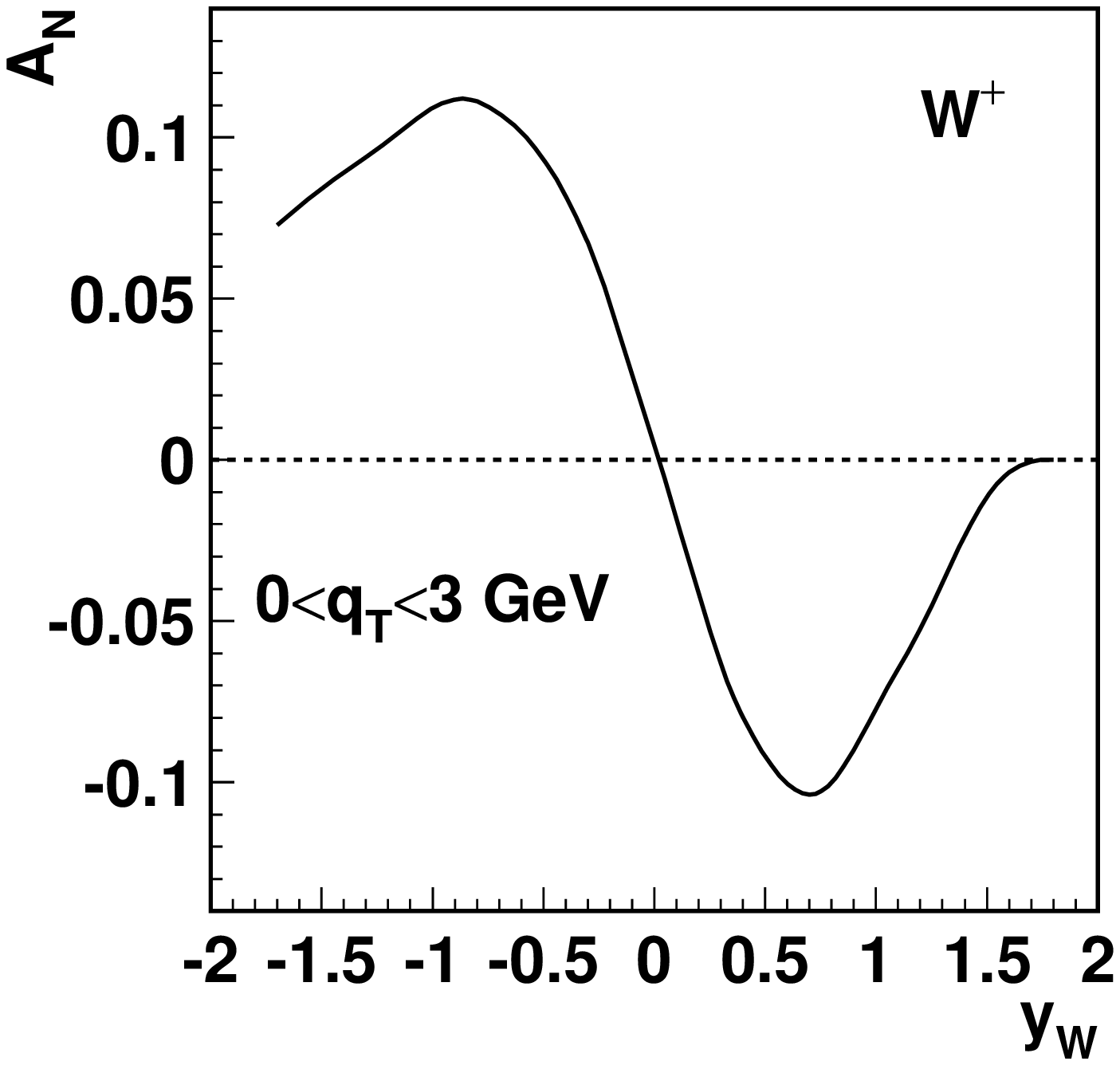,width=1.55in}
\caption{$A_N$ as a function of $W$-boson rapidity.}
\label{fig1}
\eef
\bef
\psfig{file=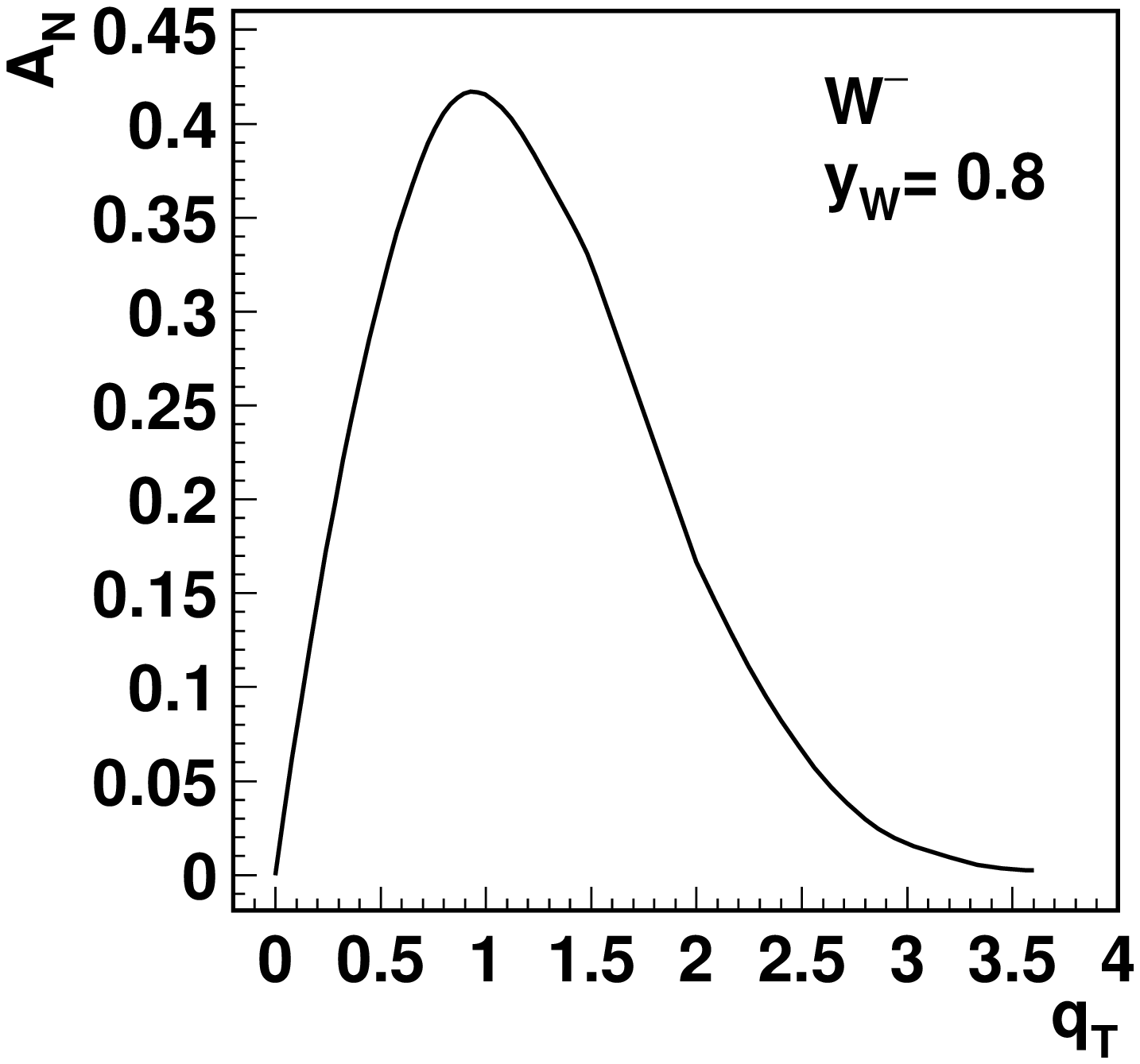,width=1.55in}
\hskip 0.1in
\psfig{file=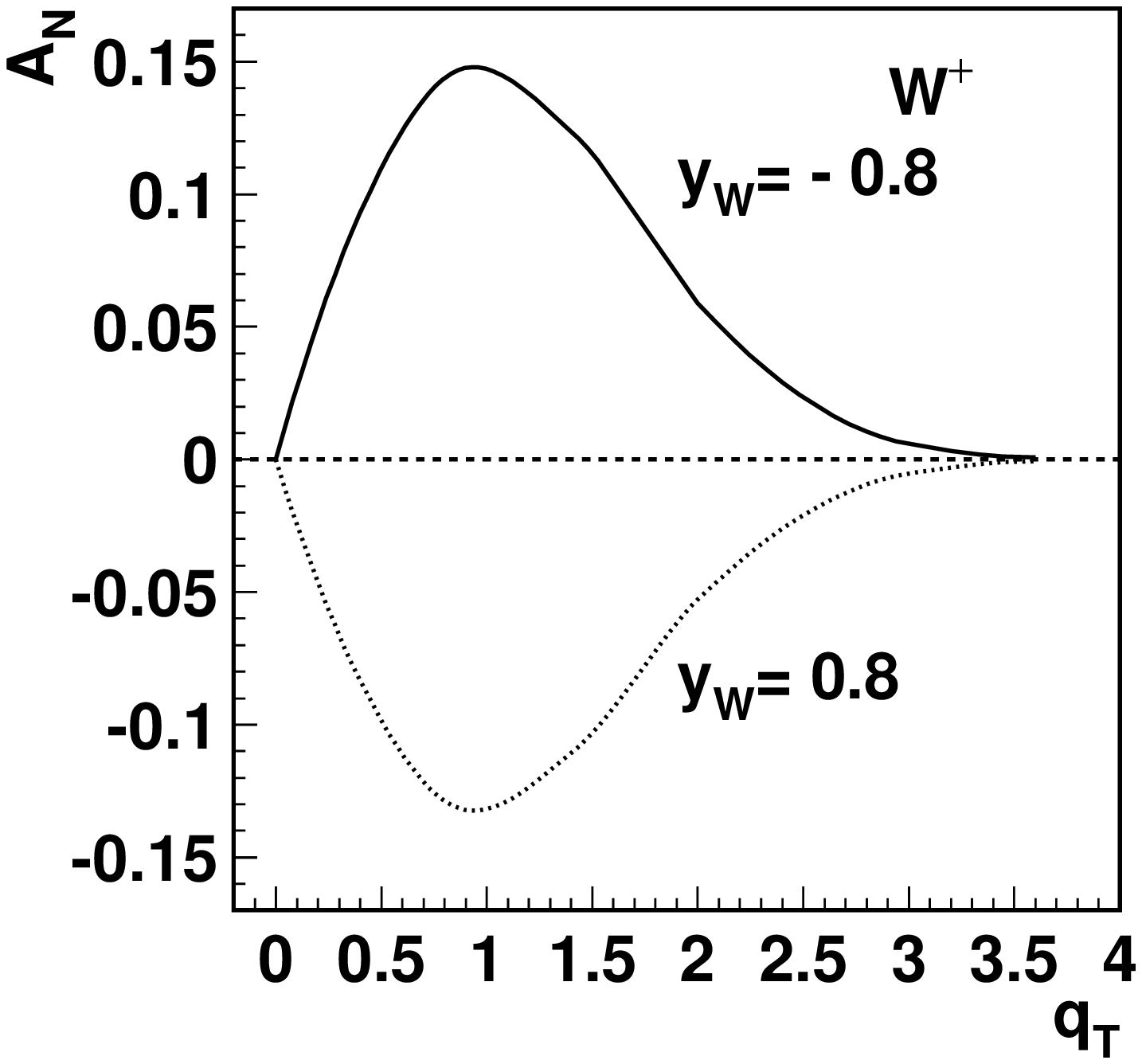,width=1.55in}
\caption{$A_N$ as a function of $W$-boson transverse momentum.}
\label{fig2}
\eef

In Figs.~\ref{fig1} and \ref{fig2}, we 
plot the $A_N$ from Eq.~(\ref{An_W}) at $\sqrt{s}=500$~GeV. 
The $W$ asymmetry is peaked at $q_T\ll M_W$ and 
is much larger than that of DY production \cite{RHIC-dy}.  
This is because the $u$ and $d$ 
Sivers functions have an opposite sign, and 
they partially cancel each other in their contribution to 
the DY asymmetry, while 
they contribute to the $W^+$ and $W^-$ separately. 
The large $W^-$ asymmetry is
caused by a large $d$ Sivers function 
\cite{Anselmino:2008sga}.  The negative $d$ Sivers
function in SIDIS gives the positive $W^-$ asymmetry.
The rapidity dependence in Fig.~\ref{fig1}
provides excellent informations for the flavor 
separation as well as the functional form 
of the Sivers function if we could reconstruct the $W$ bosons.

After integrating over the momentum of (anti)neutrino from the $W$ decay,
we obtain the leading order factorized cross section for the production of leptons of rapidity $y$ and transverse momentum $\mathbf{p}_\perp$,
\ben
\frac{d\sigma_{A^\uparrow B\to \ell(p)}(\vec{S}_\perp)}
     {dy\,d^2{\mathbf{p}_\perp}}
&=&
\sum_{a,b}\left|V_{ab}\right|^2
\int dx_a\, d^2\mathbf{k}_{a\perp} 
\int dx_b\, d^2\mathbf{k}_{b\perp} 
\nonumber \\
&\times &
f_{a/A^\uparrow}^{\rm DY}(x_a,\mathbf{k}_{a\perp},\vec{S}_\perp)\,
f_{b/B}(x_b,{k}_{b\perp})\, 
\nonumber \\
&\times &
\frac{1}{16\pi^2\hat{s}}
\left|\overline{\cal M}_{ab\to \ell}\right|^2
\delta(\hat{s}+\hat{t}+\hat{u})\, ,
\label{fac_l}
\een
where $\hat{s}$, $\hat{t}$, and $\hat{u}$ are the 
Mandelstam vaeriables and the leading order 
partonic scattering amplitude square, 
$\left|\overline{\cal M}_{ab\to\ell}\right|^2$, 
is given by 
\ben
\frac{8({\rm G}_F M_W^2)^2}{3}
\frac{\hat{u}^2}
     {(\hat{s}-M_W^2)^2+M_W^2\, \Gamma_W^2}\, 
\label{matrix_sq}
\een
for partonic channels $ab=d\bar{u},s\bar{u},\bar{d}u,\bar{s}u$;
or by the same one with the $\hat{u}^2$ replaced by $\hat{t}^2$
for the rest light flavor channels 
$ab=\bar{u}d,\bar{u}s,u\bar{d},u\bar{s}$.
$\Gamma_W$ in Eq.~(\ref{matrix_sq}) is the $W$ leptonic decay width. 
Substituting Eq.~(\ref{TMDPDF}) into Eq.~(\ref{fac_l}), we 
derive both the spin-averaged and spin-dependent cross sections, 
from which we evaluate the SSAs of inclusive lepton production
from $W$ decay numerically.

\bef
\psfig{file=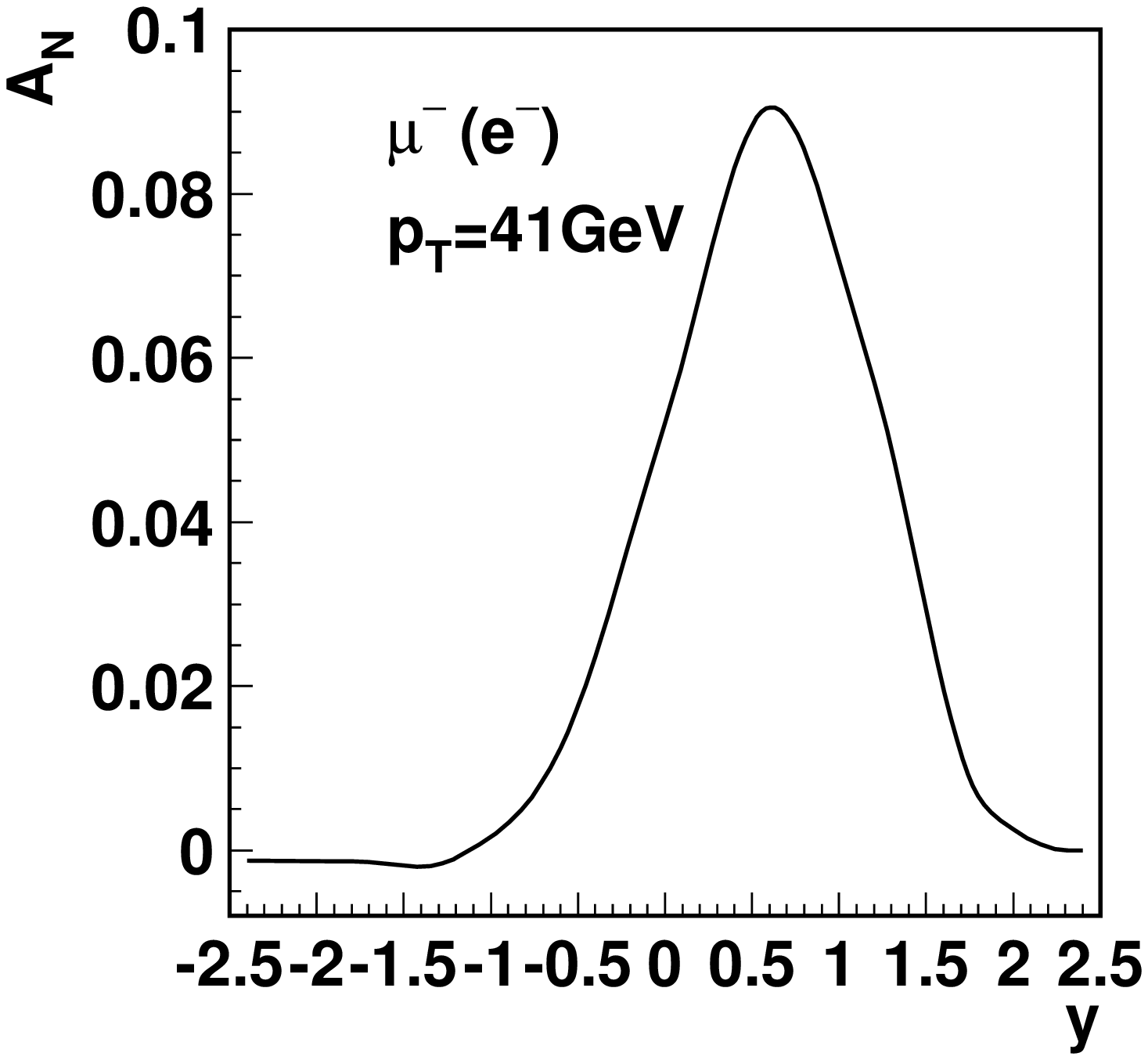,width=1.55in}
\hskip 0.1in
\psfig{file=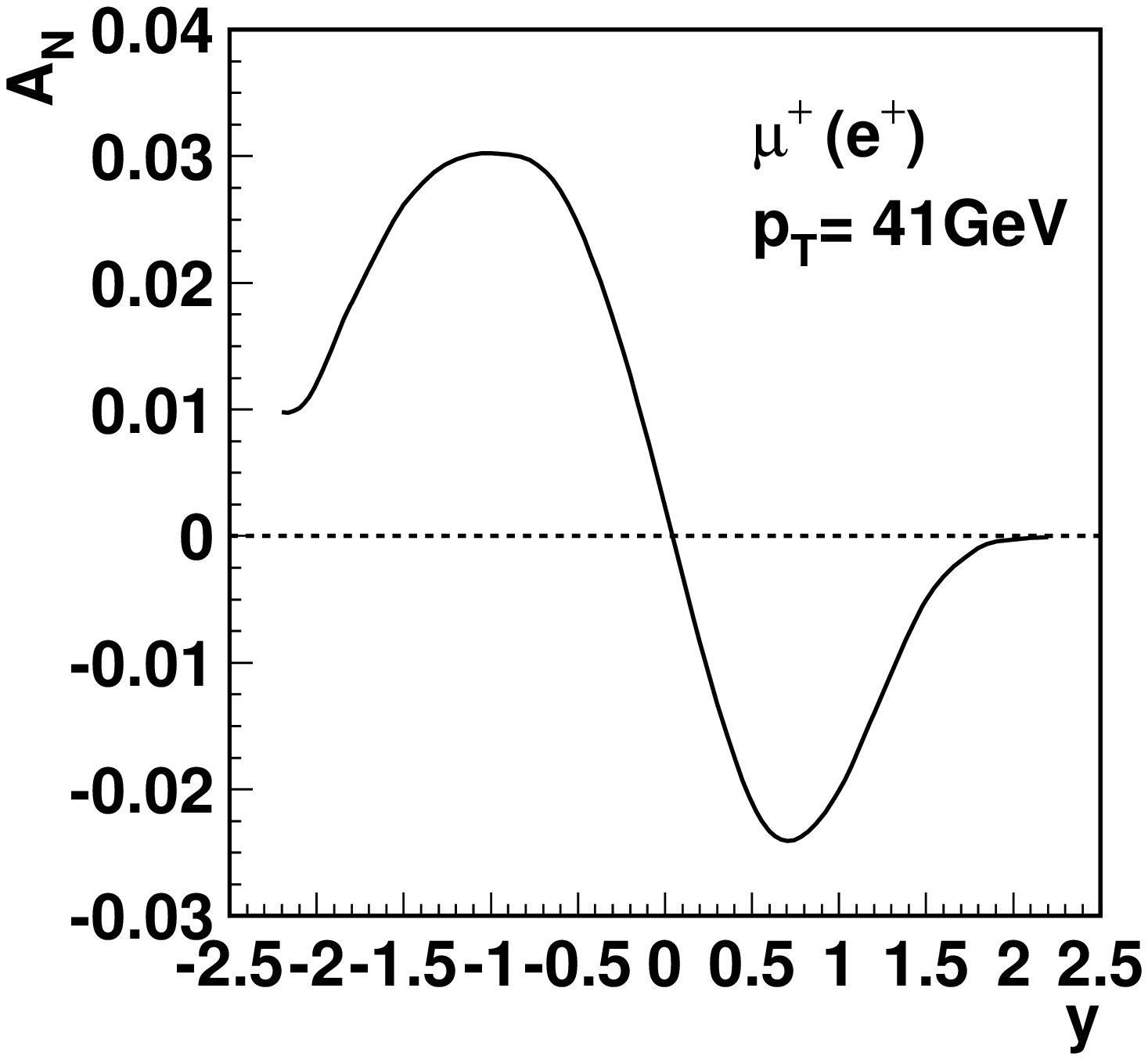,width=1.55in}
\caption{$A_N$ as a function of lepton rapidity.}
\label{fig3}
\eef
\bef
\psfig{file=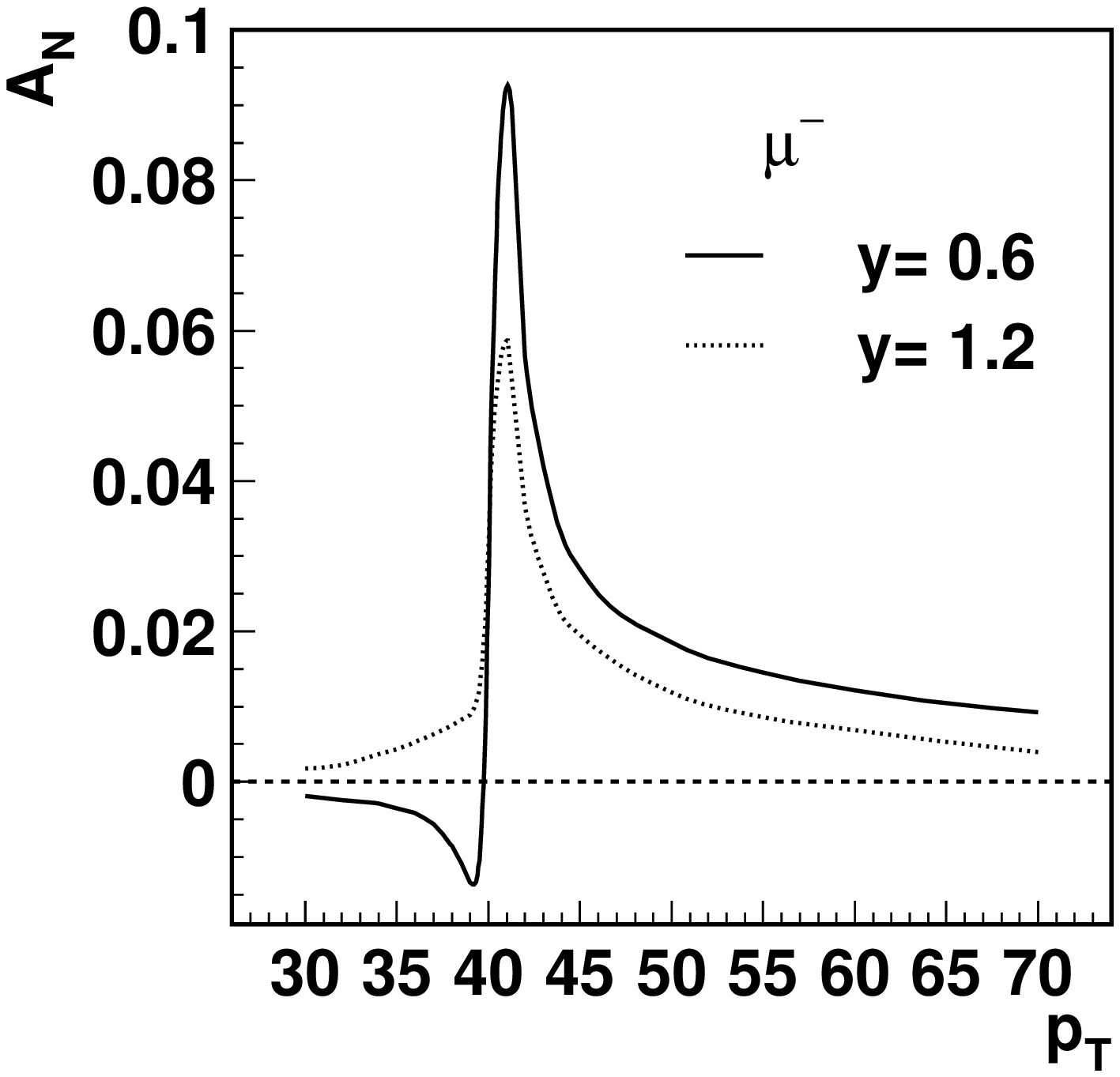,width=1.55in}
\hskip 0.1in
\psfig{file=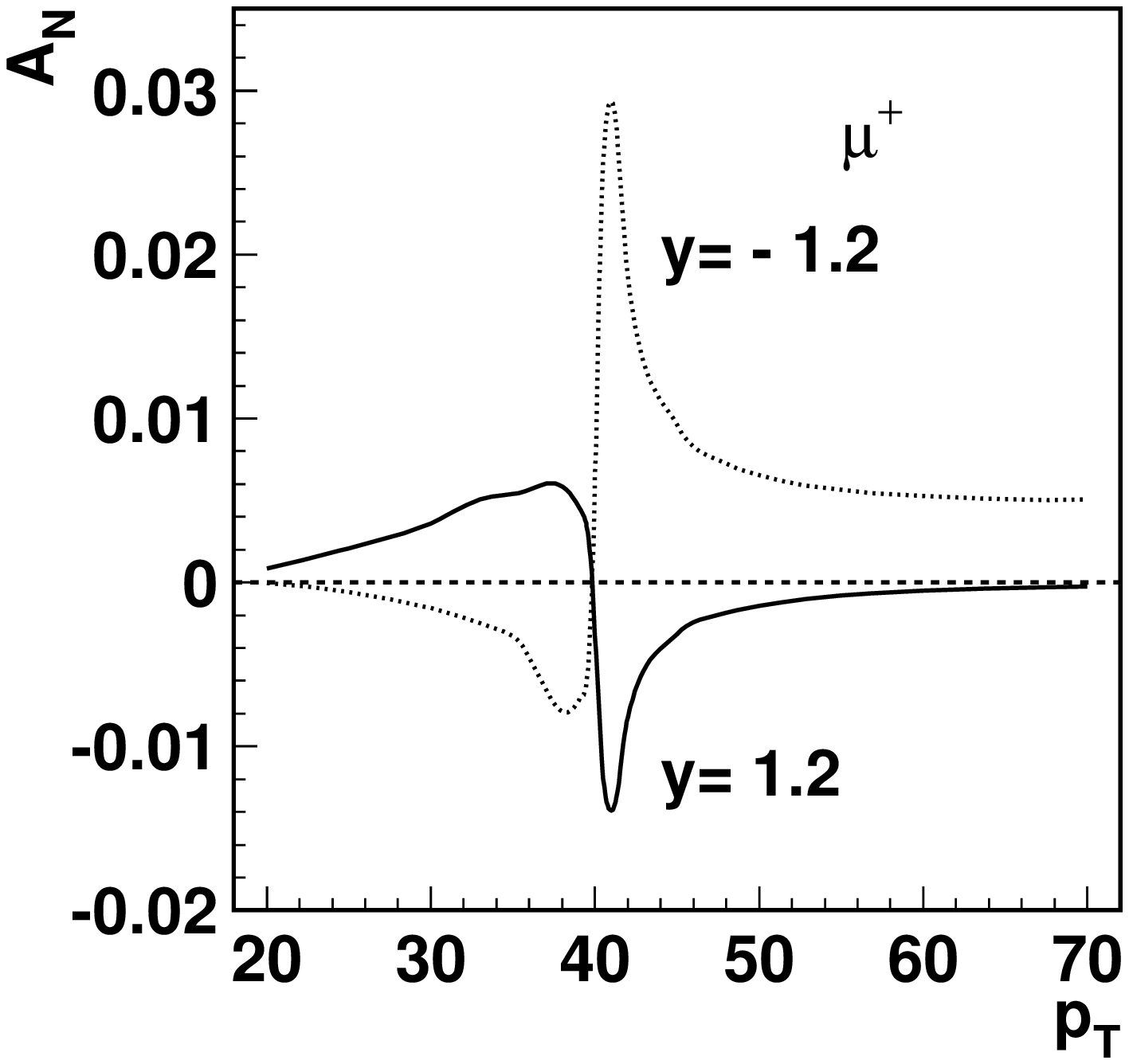,width=1.55in}
\caption{$A_N$ as a function of lepton transverse momentum.}
\label{fig4}
\eef

In Figs.~\ref{fig3} and \ref{fig4}, we present our predictions
for the inclusive lepton asymmetry from the decay of $W$ bosons
at RHIC energy.  Although the decay diluted the size of the 
asymmetry, the lepton inherited all key features of the 
$W$ asymmetry in Figs.~\ref{fig1} and \ref{fig2}.  
As shown in Fig.~\ref{fig4}, 
the lepton asymmetry is sharply peaked at $p_T\sim 41$~GeV, 
which should help control the potential background.  
The difference in rapidity dependence
of the $W^+$ and $W^-$ in Fig.~\ref{fig3} provides the excellent
flavor separation of the Sivers function, as well as rich 
information on the functional form.
For a good range of rapidity, the lepton asymmetry is measurable 
at RHIC.

{\it IV.\ Summary and Conclusions. }\
In summary, we have derived the time-reversal modified universality 
for both quark and gluon Sivers functions from the parity and 
time-reversal invariance of the gauge invariant matrix elements 
that define the TMD parton distributions.  We confirm the 
Collins' prediction for the sign change of the quark Sivers function 
in SIDIS and in DY \cite{Collins:2002kn}.  The sign change of 
the Sivers function in SIDIS and in DY is a natural property of 
the gauge invariant TMD parton distributions in QCD.  
Corresponding sign change of the SSAs, if they could be factorized 
in terms of these TMD parton distributions, is a fundamental 
prediction of QCD.

We have calculated, in terms of the TMD parton distributions, 
the SSAs of $W$ production as well as inclusive lepton production 
from the decay of $W$ bosons in polarized proton-proton collision 
at RHIC energy.  We find that although the asymmetry is 
diluted from the $W$ decay, the lepton asymmetry is at the level 
of several percent and measurable 
for a good range of lepton rapidity at RHIC.  
Because the lepton asymmetry is sharply peaked
at the $p_T\sim 41$~GeV, the potential background could be strongly 
suppressed.  We conclude that this measurable lepton asymmetry at 
high $p_T$ at RHIC is an excellent observable for measuring the 
Sivers functions of different flavors and for testing the 
time-reversal modified universality of the Sivers function.

We thank J.~Lajoie and F.~Wei for helpful discussions. 
This work was supported in part by the U. S. Department of Energy 
under Grant No.~DE-FG02-87ER40371.



\begin{thebibliography}{99}

\bibitem{CSS-fac}
  for reviews, see:
  J.~C.~Collins, D.~E.~Soper and G.~Sterman,
  Adv.\ Ser.\ Direct.\ High Energy Phys.\  {\bf 5}, 1 (1988).

\bibitem{PDFs}
  for reviews, see:
  J.~Pumplin, D.~R.~Stump, J.~Huston, H.~L.~Lai, P.~Nadolsky and W.~K.~Tung,
  JHEP {\bf 0207}, 012 (2002);
  A.~D.~Martin, R.~G.~Roberts, W.~J.~Stirling and R.~S.~Thorne,
  Eur.\ Phys.\ J.\  C {\bf 23}, 73 (2002).

\bibitem{Bunce:1976yb}
  G.~Bunce {\it et al.},
  Phys.\ Rev.\ Lett.\  {\bf 36}, 1113 (1976).

\bibitem{ssa_review}
  for reviews, see:
  U.~D'Alesio and F.~Murgia,
  Prog.\ Part.\ Nucl.\ Phys.\  {\bf 61}, 394 (2008).

\bibitem{Kane:1978nd}
  G.~L.~Kane, J.~Pumplin and W.~Repko,
  Phys.\ Rev.\ Lett.\  {\bf 41}, 1689 (1978).

\bibitem{QS_fac}
  J.~W.~Qiu and G.~Sterman,
  AIP Conf.\ Proc.\  {\bf 223}, 249 (1991);
  Nucl.\ Phys.\  B {\bf 353}, 137 (1991).

\bibitem{Efremov}
  A.~V.~Efremov and O.~V.~Teryaev,
  Sov.\ J.\ Nucl.\ Phys.\  {\bf 36}, 140 (1982).
  Phys.\ Lett.\ B {\bf 150}, 383 (1985).

\bibitem{qiu}
J.~W.~Qiu and G.~Sterman,
Phys.\ Rev.\ Lett.\  {\bf 67}, 2264 (1991),
  Nucl.\ Phys.\ B {\bf 378}, 52 (1992).

\bibitem{Kang:2008ey}
  for example, see:
  Z.~B.~Kang and J.~W.~Qiu,
  Phys.\ Rev.\  D {\bf 79}, 016003 (2009),
  and references therein.

\bibitem{Collins:2002kn}
  J.~C.~Collins,
  Phys.\ Lett.\  B {\bf 536}, 43 (2002).

\bibitem{Boer:2003cm}
  D.~Boer, P.~J.~Mulders and F.~Pijlman,
  Nucl.\ Phys.\  B {\bf 667}, 201 (2003).

\bibitem{Ji:2004xq}
  X.~d.~Ji, J.~P.~Ma and F.~Yuan,
  Phys.\ Lett.\  B {\bf 597}, 299 (2004),
  Phys.\ Rev.\  D {\bf 71}, 034005 (2005).

\bibitem{Collins:2004nx}
  J.~C.~Collins and A.~Metz,
  Phys.\ Rev.\ Lett.\  {\bf 93}, 252001 (2004).

\bibitem{Collins:2007nk}
  J.~Collins and J.~W.~Qiu,
  Phys.\ Rev.\  D {\bf 75}, 114014 (2007).

\bibitem{Sivers}
D.~W.~Sivers,
Phys.\ Rev.\ D {\bf 41}, 83 (1990),
{\bf 43}, 261 (1991).

\bibitem{Collins}
  J.~C.~Collins,
  Nucl.\ Phys.\  B {\bf 396}, 161 (1993).

\bibitem{UnifySSA}
  X.~Ji, J.~W.~Qiu, W.~Vogelsang and F.~Yuan,
  Phys.\ Rev.\ Lett.\  {\bf 97}, 082002 (2006),
  Phys.\ Rev.\  D {\bf 73}, 094017 (2006),
  Phys.\ Lett.\  B {\bf 638}, 178 (2006);
  Y.~Koike, W.~Vogelsang and F.~Yuan,
  Phys.\ Lett.\  B {\bf 659}, 878 (2008).

\bibitem{Anselmino:2008sga}
  M.~Anselmino {\it et al.},
  Eur.\ Phys.\ J.\  A {\bf 39}, 89 (2009)

\bibitem{RHIC-dy}
  J.~C.~Collins {\it et al.},
  Phys.\ Rev.\  D {\bf 73}, 094023 (2006).

\bibitem{tmd_gauge}  
  X.~d.~Ji and F.~Yuan,
  Phys.\ Lett.\  B {\bf 543}, 66 (2002);
  A.~V.~Belitsky, X.~Ji and F.~Yuan,
  Nucl.\ Phys.\  B {\bf 656}, 165 (2003).
  
\bibitem{ssa_W}
  S.~J.~Brodsky, D.~S.~Hwang and I.~Schmidt,
  Phys.\ Lett.\  B {\bf 553}, 223 (2003);
  I.~Schmidt and J.~Soffer,
  Phys.\ Lett.\  B {\bf 563}, 179 (2003).

\bibitem{GRV98}
  M.~Gluck, E.~Reya and A.~Vogt,
  Eur.\ Phys.\ J.\  C {\bf 5}, 461 (1998).
  
\end{thebibliography}
\end{document}